\documentclass{emulateapj}

\def \arcmin {\hbox{$^\prime$}}
\def \arcsec {\hbox{$^{\prime\prime}$}}

\def\spose#1{\hbox to 0pt{#1\hss}}
\def\ltsim{$\mathrel{\spose{\lower 3pt\hbox{$\sim$}}
        \raise 2.0pt\hbox{$<$}}$\thinspace}
\def\gtsim{$\mathrel{\spose{\lower 3pt\hbox{$\sim$}}
        \raise 2.0pt\hbox{$>$}}$\thinspace}
\def \msun {${\rm M_\odot}$}

\newcommand\solar{\hbox{{$Z_{\odot}$}}}

\newcommand{\source}{\mbox{XMMU\,J131359.7--162735}}

\newcommand{\apec}{APEC}

\newcommand{\xspec }{{\em Xspec}}

\newcommand{\fxunits}{\mbox{ergs cm$^{-2}$ s$^{-1}$}}
\newcommand{\lxunits}{\mbox{ergs s$^{-1}$}}
\newcommand{\xmm }{{\em XMM}}

\newcommand{\rosat }{{\em ROSAT}}

\newcommand\omegam{\hbox{{$\Omega_{\rm m}$}}}
\newcommand\omegalambda{\hbox{{$\Omega_{\Lambda}$}}}
\newcommand\kmsmpc{{\rm km s$^{-1}$ Mpc$^{-1}$}}
\newcommand\ho{\hbox{{$H_{0}$}}}
\slugcomment{Submitted Nov 27, 2006; Accepted March 16, 2007 for publication in the \rm{Astrophysical Journal}}
\shorttitle{\xmm\ discovery of a cluster of galaxies at $\MakeLowercase{z}=0.28$}
\shortauthors{Gastaldello et~al.}

\begin{document}

\title{Serendipitous \emph{XMM-Newton} discovery of a cluster of galaxies at
$\MakeLowercase{z}=0.28$}

\author {Fabio Gastaldello\altaffilmark{1},
         David A. Buote\altaffilmark{1},
         Philip J. Humphrey\altaffilmark{1}, 
         Luca Zappacosta\altaffilmark{1},
         Marc S. Seigar\altaffilmark{1},
         Aaron J. Barth\altaffilmark{1},  
         Fabrizio Brighenti\altaffilmark{2,3}, 
         \& William G. Mathews\altaffilmark{2}
}
\altaffiltext{1}{Department of Physics and Astronomy, University of
California at Irvine, 4129
Frederick Reines Hall, Irvine, CA 92697-4575}
\altaffiltext{2}{UCO/Lick Observatory, University of California at Santa Cruz,
 1156 High Street, Santa Cruz, CA 95064}
\altaffiltext{3}{Dipartimento di Astronomia, Universit\`a di Bologna, via
Ranzani 1, Bologna 40127, Italy}
\begin{abstract}
We report the discovery of a galaxy cluster serendipitously 
detected as an extended X-ray source in an offset observation of 
the group NGC 5044. The cluster redshift, $z$=0.281, determined from the 
optical spectrum of the brightest cluster galaxy, agrees with that inferred 
from the X-ray spectrum using the Fe K$\alpha$ complex of the hot ICM 
($z=0.27\pm0.01$).
Based on the 50\,ks \xmm~observation, we find that within a radius of 383 
kpc the cluster has an unabsorbed \mbox{X-ray} flux,
$f_{\rm X}\,(0.5-2\,{\rm{keV}})=(3.34^{+0.08}_{-0.13}) \times 10^{-13}$ 
\fxunits, a bolometric \mbox{X-ray} luminosity, 
$L_{\rm X}=(2.21^{+0.34}_{-0.19}) \times 10^{44}$ \lxunits, 
$kT=3.57\pm0.12$\,keV, and metallicity, $0.60\pm0.09$ \solar.
The cluster obeys the scaling relations for $L_{\rm X}$ and $T$ observed at 
intermediate redshift.  
The mass derived from an isothermal NFW model fit is, 
$M_{\rm vir} = 3.89\pm0.35 \times10^{14}$ \msun, with a concentration 
parameter, $c = 6.7\pm0.4$, consistent with the range of values
expected in the concordance cosmological model for relaxed clusters.
The optical properties suggest this could be a ``fossil cluster''.
\end{abstract}


\section{Introduction} 
\label{Introduction} 

Galaxy groups and low mass clusters ($T < 4$ keV) are 
starting to be detected and analyzed in detail in the X-ray band 
at intermediate redshift $0.2 < z < 0.6$ \citep[e.g.,][]{willis05,jeltema06}. 
These objects are more likely to display the effects of 
non-gravitational energy injection into the intracluster medium (ICM) than 
hotter, more massive, clusters \citep[e.g.,][]{voit05b}. 
The study of extended X-ray
objects over a broad temperature range at $z> 0.2$ will provide  
important insight into the evolution of their hot gas and the 
deviation of X-ray scaling relations from simple, self-similar expectations.
Studies of objects in the redshift range $0.2 < z < 0.6$ with 
$kT \sim 2-3$ keV already suggest that they are less 
dynamically evolved than their counterparts at $z=0$ \citep{mulch06}.

The combination of surveys specifically designed to detect clusters 
\citep[e.g. the \xmm-LSS,][]{pierre04}, with serendipitous observations, 
which can make use of deeper exposures, will further advance the knowledge in 
this field.
Here we present the discovery of a cluster at the redshift of $z=0.281$, 
serendipitously observed during an offset observation of the nearby group 
NGC 5044. All distance-dependent quantities have been computed assuming 
\ho = 70 \kmsmpc, \omegam = 0.3 and \omegalambda = 0.7. 
At the redshift of $z=0.281$, 1\arcmin\ corresponds to 255 kpc. All the errors 
quoted are at the 68\% confidence limit.

\section{\mbox{X-ray} Analysis}
\label{x-ray}

\begin{figure*}[t]
\parbox{0.5\textwidth}{
\centerline{\includegraphics[width=0.35\textwidth]{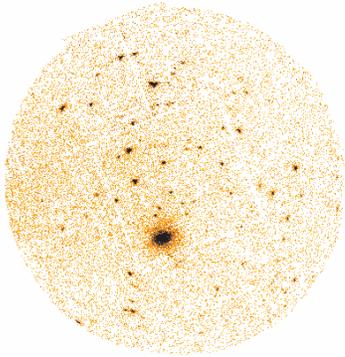}}
}
\parbox{0.5\textwidth}{
\centerline{\includegraphics[width=0.4\textwidth]{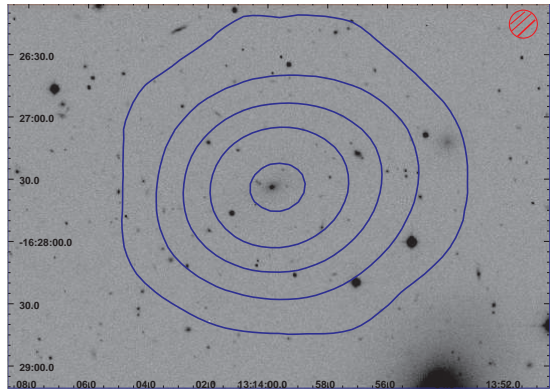}}
}
\caption{\label{fig.1} \footnotesize
{\em Left:} Exposure corrected 0.5-2.0 keV combined MOS1 and MOS 2 X-ray 
image of the observed offset field of NGC 5044. The cluster \source\ 
is clearly visible as the extended source. 
{\em Right:} The $V$ band image 
with the smoothed X-ray contours overlaid and logarithmically spaced between 2 and 37 
cts/pixel. Coordinates are J2000. The red circle shows the FWHM (6.7\arcsec) of the \xmm\ MOS PSF at the off-axis angle of the source. }
\end{figure*}

The object \source\ was detected in an offset observation of 
the group NGC 5044 (obsID 0301290101) performed on 6 January 2006. 
The source (see Fig.\ref{fig.1}) is located at 4.7\arcmin\ off-axis, and it is 
clearly extended on a $\sim2$ arc-minute scale. The X-ray centroid of 
\source\ in equatorial coordinates is 
\mbox{$\alpha_{\rm J2000.0}=13^{\rm h}13^{\rm
m}59.7^{\rm s},\, \delta_{\rm J2000.0}=-16^{\circ}27^{\rm m}35^{\rm s}$}.

The data were reduced with SAS v7.0.0 using the tasks
{\em emchain} and {\em epchain}. We considered only event patterns
0-12 for MOS and 0 for pn. We cleaned the data using the standard 
procedures: bright pixels and hot columns were removed by applying the 
expression FLAG == 0, and we corrected for pn out-of-time events.
Periods of high background due to soft protons were filtered as in 
\citet{gast06a}; the observation was very mildly affected by flares
leading to only $\sim 5$ ks of lost data, with net exposures of
52, 50 and 45 ks respectively for MOS1, MOS2 and pn.

\begin{figure}[h]
\begin{center} 
\includegraphics[width=0.3\textwidth,angle=-90]{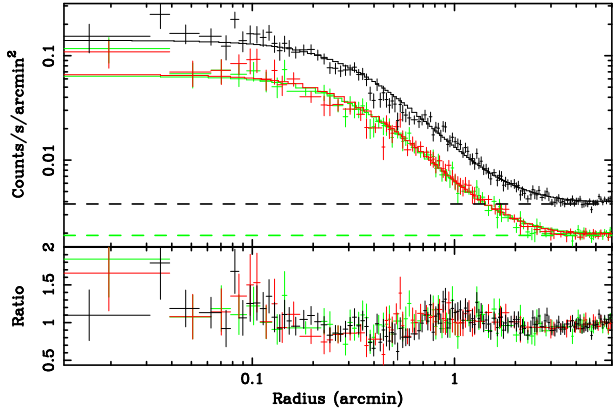}
\vskip 0.5cm
\caption{Surface brightness profile of the X-ray emission of \source. Data from MOS1, MOS2 and pn are plotted in green, red and black respectively. The best fit beta model and ratio of data over the model are also shown. The background level for PN and for one of the MOS are shown as black and green dashed lines respectively.}
\label{fig.2} \end{center}
\end{figure}

For each detector we created images in the 0.5-2 keV band with point sources 
masked out using circular regions of 25\arcsec\ radius centered at each source 
position. The images have been exposure-corrected, and a radial surface 
brightness profile was extracted 
from a circular region of 6\arcmin\ radius positioned at the cluster 
centroid. We account for the X-ray background and the emission of the NGC 5044
group by including a constant background component. 
The data were grouped to 
have at least 20 counts per bin in order to apply the $\chi^{2}$ statistic.
The fitted model is folded with the \xmm\ PSF at energy of 1 keV. The joint 
best-fit $\beta$ model \citep{beta} has a core radius of 
$r_c = 66\pm6$ kpc (15.5\arcsec$\pm1.4$\arcsec) and $\beta=0.54\pm0.01$ for 
$\chi^{2}$/d.o.f. = 465/273. Although formally 
unacceptable, the surface brightness model well-describes the overall shape 
of the profile, as evinced by the low fractional residuals (see 
Fig.\ref{fig.2}).
Fits to the profiles of the individual 
detectors give consistent results within $1\sigma$ of the combined-fit result.
If we use larger extraction regions, the emission by NGC 5044 can no longer 
be treated as constant: if we account for the emission of NGC 5044 by adding 
a second $\beta$ model at an offset of 
20.7\arcmin\ (the distance between the source and the emission peak of 
NGC 5044) with $r_c = 48\arcsec$ and $\beta = 0.53$ \citep{buot03a}, the 
resulting parameters for the source are consistent within the $1\sigma$ errors.

We extracted spectra for each detector from a 1.5\arcmin\
circular region positioned at the centroid of the emission, chosen to 
maximize the S/N over the background. 
(Choosing an extraction radius of 2\arcmin\ does not change the spectral 
parameters quoted below). Redistribution matrix files (RMFs) and
ancillary response files (ARFs) were generated using the SAS tasks
{\em rmfgen} and {\em arfgen}, the latter in extended source
mode. Appropriate flux-weighting was performed for RMFs, using our own
dedicated software, and for ARFs, using exposure-corrected images of
the source as detector maps (with pixel size of 1\arcmin, the minimum
scale modeled by {\em arfgen}) to sample the variation in emission,
following the prescription of \citet{saxton02}. The background was estimated 
locally using spectra extracted from a 2\arcmin-3\arcmin\ annular region 
positioned at the centroid of the emission; using a 3\arcmin-4\arcmin\
annulus gives results consistent within their $1\sigma$ errors.
We checked that vignetting corrections for the sky background components do 
not affect the results by exploring the complementary approach of a 
complete modeling of the background components (including the thermal 
emission of NGC 5044) as in \citet{gast06a}: the spectral results obtained 
are consistent with the ones quoted below within their $1\sigma$ errors.

\begin{figure}[t]
\begin{center} 
\includegraphics[width=0.3\textwidth,angle=-90]{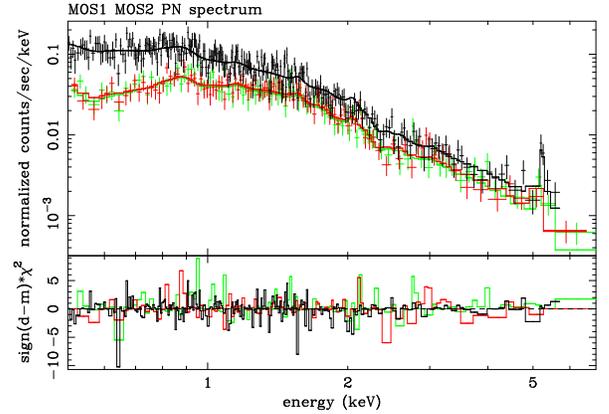} 
\caption{X-ray spectrum of the source taken from a 1.5\arcmin\ aperture positioned at the centroid of the emission. Data from MOS1, MOS2 and pn are plotted in green, red and black respectively. The best fit model and residuals are also shown.}
\label{fig.3} \end{center}
\end{figure}

The spectra from the three detectors were re-binned to ensure a signal-to-noise
ratio of at least 3 and a minimum 20 counts per bin, and they were 
jointly fitted with an \apec\ thermal plasma modified by Galactic 
absorption \citep{dick90} . The spectral fitting was performed with 
\xspec\ \citep[ver11.3.1,][]{xspec} and quoted metallicities are relative to 
the abundances of \citet{grsa98}. We display the spectra in Fig.\ref{fig.3}.
The source photons correspond to about 77\% of the total events ($\sim 4400$ 
counts in each MOS and $\sim7700$ in the pn). An emission line is clearly 
visible corresponding to the Fe K$\alpha$ complex emitted at a redshift of 
$z=0.27\pm0.01$.
The redshift measured is in close agreement with the redshift of $z=0.2814$ 
determined optically from the spectrum of the brightest cluster galaxy 
(see \S\ref{optical}). Fixing the redshift value to the optical determination 
gives as best fit parameters, $T = 3.57\pm0.12$ keV, and,  $Z=0.60\pm0.09$ 
\solar, for $\chi^{2}$/d.o.f. = 486/463. 

Using the best-fit model, the unabsorbed flux within the aperture of 
radius 1.5\arcmin\ (383 kpc) is 
$3.34^{+0.08}_{-0.13}\times10^{-13}$ \fxunits\ in 
the 0.5-2 keV band. This corresponds to an unabsorbed luminosity of 
$7.57^{+0.10}_{-0.40}\times10^{43}$ \lxunits\ in the 0.5-2 keV band and to a 
bolometric (0.01-100 keV) luminosity of $2.21^{+0.34}_{-0.19}\times10^{44}$ 
\lxunits. The quoted errors on flux and luminosity are obtained by \xspec\ 
using a Montecarlo procedure.

To investigate possible spatial variation in the spectral 
parameters of the cluster, we extracted two annular regions of radii 
0\arcmin-0.5\arcmin\ and 0.5\arcmin-1.5\arcmin. The derived spectral 
parameters are: $T=3.64\pm0.14$ keV and $Z=0.66\pm0.13$ \solar\ with 
$\chi^{2}$/d.o.f. = 280/231 for the inner annulus; $T=3.48^{+0.19}_{-0.16}$ 
keV and $Z=0.52\pm0.13$ \solar\ with $\chi^{2}$/d.o.f. = 280/271 for 
the outer annulus. The width of the bins has been chosen in order to avoid 
bias in the temperature measurements caused by scattered flux by the PSF 
( 80\% encircled energy fraction radius is 25\arcsec\ at 1.5 keV and at the 
off-axis angle of the source). 
The cluster is therefore consistent with being isothermal over the explored 
radial range.

\source\ was also serendipitously detected in the \rosat\ PSPC pointed 
observation of NGC 5044 \citep[1WGA J1313.9-1627;][]{white94}, though it was 
very close to the detector gaps. This is likely the reason why with \rosat\ 
the source appeared not to be extended and had a lower flux 
($1.75\pm0.14 \times 10^{-13}$ \fxunits\ in the 0.2-2 keV band).

The cluster has regular X-ray isophotes and is centered on a dominant early 
type galaxy (see \S\ref{optical}). These characteristics suggest the cluster 
is relaxed and that hydrostatic equilibrium is a good approximation.
We calculated the total mass profile using two different models.
First, we used the best-fit $\beta$ model (see \S \ref{x-ray}) for which 
the gas density and total mass profiles can be expressed by simple 
analytical formula \citep[e.g.,][]{ettori00}. 
We evaluated $r_{500}$ as the radius at which the density is 500 times the 
the critical density and the virial radius as the radius at which the
density corresponds to $\Delta_{\rm{vir}}$, as obtained by \citet{bryan98} for 
the concordance cosmological model used in this paper.
To evaluate the errors on the estimated quantities,
we repeat the measurements after 10000 random selections of a temperature and
parameters of the surface brightness profile, which were drawn from Gaussian 
distributions with mean and variance in accordance with the best-fit results.
For $\Delta=500$ we obtained, 
$M_{500} = (1.60\pm0.09) \times 10^{14}$ \msun\ within 
$r_{500} = 750\pm14$ kpc; 
the virial mass is, $M_{\rm{vir}}= (3.24\pm0.18)\times10^{14}$ \msun, 
within the virial radius $r_{\rm{vir}} = 1511\pm29$ kpc. 
Secondly, we fit the surface brightness profile 
with an isothermal NFW model \citep{suto98}. 
We obtain a concentration parameter, 
$c=6.7\pm0.4$, virial radius $r_{\rm{vir}} = 1607^{+48}_{-42}$ kpc,
and virial mass $M_{\rm{vir}}=(3.89\pm0.35) \times10^{14}$ \msun, with  
$\chi^{2}$/d.o.f. = 424/273. The mass determinations using the two different 
models are in agreement within the $1\sigma$ errors.

\section{Optical Follow-up Observations}
\label{optical}

A $V$-band and $I$-band image of the field were taken with the direct
imaging CCD on the 2.5-m du Pont Telescope at Las Campanas Observatory on 
the night of 27 January 2006.
In both wavebands, 2$\times$300 s exposures were taken in clear, photometric 
conditions.
Photometric zero-points and extinction corrections were determined by
observing three standard stars chosen from the \citet{landolt92} catalog. 
The images have been processed and cleaned of CCD defects and cosmic rays. 
The $V$-band image 
is shown with X-ray contours superposed in Fig.\ref{fig.1}. The image 
reveals an over-density of galaxies and, in particular, a bright dominant 
galaxy at the same location as the X-ray centroid. 

\begin{figure}
\centerline{\includegraphics[width=0.26\textwidth,angle=-90]{f4a.ps}}
\centerline{\includegraphics[width=0.28\textwidth,angle=-90]{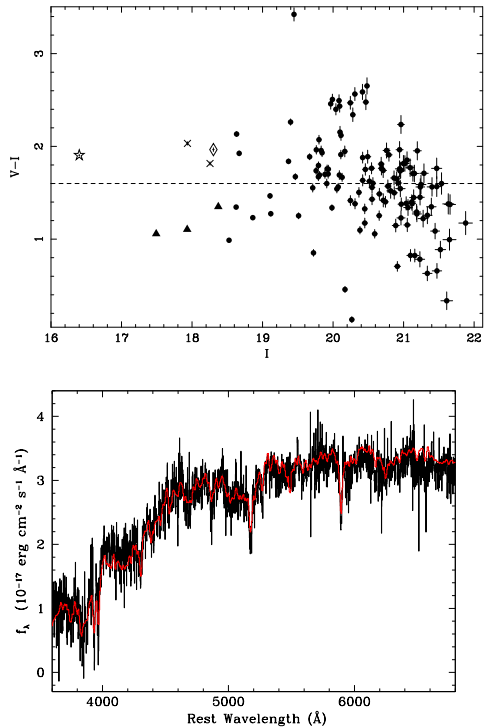}}
\caption{{\em Top:} 
Color-magnitude diagram of the galaxies within 2.8\arcmin\ of \source.
The horizontal dashed line indicates the predicted color of a $z=0.28$ cluster 
elliptical. The star indicates the BCG. Triangles indicate likely dwarf 
members of NGC 5044 
\citep[one of them is the object \#26 in the catalog of][]{ferguson90}.
Crosses indicate objects very far from the BCG (see text). The diamond 
indicates the only other object within 2 magnitudes of the BGC.
{\em Bottom:} 
A portion of the Keck ESI spectrum, binned to 2
\AA/pixel.  Over-plotted in red is the elliptical galaxy template
spectrum from the spectral atlas of \citet{kinney96}.
\label{fig.4}}
\end{figure}

Photometry was performed using SExtractor \citep{bertin96} with automatic 
aperture magnitudes. 
We plot in the top panel of 
Fig.\ref{fig.4} the color-magnitude diagram of the galaxies detected 
within 3.1\arcmin\ radius of the central galaxy (corresponding to 803 kpc, 
half of the estimated virial radius): the horizontal dashed line corresponds 
to the expected $V-I$ color of a passively evolving elliptical galaxy at 
$z\sim0.28$, which is $V-I\sim1.6$ for a formation redshift, $z>5$ 
\citep{nelson02}.
The difference in magnitude between the central 
galaxy ($m_I = 16.4$) and the likely cluster members is very close to the 2 
magnitudes of the definition of fossil systems \citep{jones03}. 
We have excluded three objects (triangles in Fig.\ref{fig.4}) because they are 
likely members of the NGC 5044 group. A further two objects, which are within 
2 magnitudes of the BCG, are  
very far from the cluster center (2.6\arcmin, 664 kpc, and  2.5\arcmin, 638 
kpc, crosses in Fig.\ref{fig.4}). 
A possible member is 1.9 magnitudes fainter 
than the BCG at a distance of 1.8\arcmin\ (diamond in Fig.\ref{fig.4}).

We obtained an optical spectrum of the central galaxy on the night of
7 March 2006 UT, using the ESI spectrograph \citep{sheinis02} at
the Keck-II telescope.  A 0\farcs75 slit width was used, giving a
spectral resolution of $\sim50$ km s$^{-1}$ (FWHM) and a pixel scale of
11.4 km s$^{-1}$ pixel$^{-1}$.  The galaxy was observed at airmass 1.25 for
900 s in partly cloudy conditions, and the slit
was oriented at the parallactic angle.  The spectrum was extracted
with a 1\arcsec\ extraction width, and wavelength calibrated and
flux-calibrated using an observation of the standard star Feige 34.
The individual echelle orders were combined into a single
spectrum. 
The galaxy redshift was measured by performing a direct fit of a K3III
stellar spectrum (observed on the same night with ESI),
velocity-broadened by convolution with a Gaussian kernel.  The
best-fitting model yields a recession velocity of $84360\pm19$ km
s$^{-1}$, or $z = 0.2814\pm0.0001$, in close agreement with the
redshift estimated from the X-ray observations.
The ESI spectrum is shown in the bottom panel of Fig.\ref{fig.4}.

\section{Discussion}
\label{discussion}

This serendipitously discovered cluster obeys the scaling relations for 
$L_{\rm X}$ and $T$ derived at 
intermediate redshift. The aperture of 383 kpc used for spectroscopy encloses 
78.3\% of the flux within $r_{500}$, assuming the cluster emission profile 
follows the $\beta$ model derived within that aperture. 
The derived bolometric luminosity within $r_{500}$ is 
$L_{500} = (2.82^{+0.43}_{-0.24}) \times 10^{44}$ \lxunits.
The derived $L_{500}$ is in good agreement with the $L-T$ relation found 
for the sample of low redshift clusters ($z<0.09$) by \citet{mark98b},  
the low-redshift groups ($z<0.03$) in the GEMS sample \citep{osmond04}, 
the six groups/poor clusters at intermediate redshift ($0.29<z<0.44$) in the 
\xmm\ LSS survey \citep{willis05}, and the six groups/poor clusters 
at intermediate redshift ($0.23<z<0.59$) in the sample of \citet{jeltema06}.

The optical appearance of the cluster suggests tantalizing evidence that this 
object may be one of the rare fossil clusters \citep[3\%-6\% of the total 
number according to][]{milos06}. More detailed optical imaging and 
spectroscopy are required to investigate this possibility.
The derived total magnitude for the BCG, corrected for Galactic extinction, 
as well as for the $k$-correction \citep{fukugita95}, is $M_I = -24.78$,
which corresponds to $M_B = -22.38$ and $M_R = -24.05$, adopting colors 
$B-I_J = 2.40$ and $R_J-I_J = 0.73$ \citep{fukugita95}. 
The BCG of \source\ is among the most luminous in the sample 
of BCGs in fossil systems of \citet{khosroshahi06}. The measured values of 
$L_B$ of the BCG and $L_X$ put the object very close to the locus of baryonic 
closure \citep{mathews05}, as expected for fossil systems.

The measured $c$, multiplied by the expected 
dependence of $1+z$ \citep{bullock01}, is consistent with a $\Lambda$CDM model 
with $\sigma_8=0.9$ indicative of a relaxed, early forming system in 
agreement with the theoretical prediction for 
fossil systems \citep{zentner05,donghia05}
and with the observational results of relaxed, low-$z$ clusters 
\citep{buote06b}.

\section{Conclusions}
\label{conclusion}
The discovery of \source\ reported in this letter represents 
the deepest observation among the intermediate redshift systems discussed in 
the literature \citep{willis05,jeltema06}, and it already provides an 
interesting description of its fundamental parameters and dynamical state. 
Future optical spectroscopy and dedicated X-ray follow-up will allow
detailed spectral pro\-per\-ties to be measured.

\begin{acknowledgements}

We thank the referee, H. Ebeling, for useful comments and suggestions.
Partial support for this work was provided by NASA-XMM grant NNG06GC48G.
The authors wish to acknowledge the very significant
cultural role and reverence that the summit of Mauna Kea has always
had within the indigenous Hawaiian community.  We are most fortunate
to have the opportunity to conduct observations from this mountain.
\end{acknowledgements}




\begin{thebibliography}{0}
\expandafter\ifx\csname natexlab\endcsname\relax\def\natexlab#1{#1}\fi
\expandafter\ifx\csname bibnamefont\endcsname\relax
  \def\bibnamefont#1{#1}\fi
\expandafter\ifx\csname bibfnamefont\endcsname\relax
  \def\bibfnamefont#1{#1}\fi
\expandafter\ifx\csname citenamefont\endcsname\relax
  \def\citenamefont#1{#1}\fi
\expandafter\ifx\csname url\endcsname\relax
  \def\url#1{\texttt{#1}}\fi
\expandafter\ifx\csname urlprefix\endcsname\relax\def\urlprefix{URL }\fi
\providecommand{\bibinfo}[2]{#2}
\providecommand{\eprint}[2][]{\url{#2}}

\end{thebibliography}


\begin{thebibliography}{32}
\expandafter\ifx\csname natexlab\endcsname\relax\def\natexlab#1{#1}\fi

\bibitem[{{Arnaud}(1996)}]{xspec}
{Arnaud}, K.~A. 1996, in ASP Conf. Ser. 101: Astronomical Data Analysis
  Software and Systems V, Vol.~5, 17

\bibitem[{{Bertin} \& {Arnouts}(1996)}]{bertin96}
{Bertin}, E. \& {Arnouts}, S. 1996, \aaps, 117, 393

\bibitem[{{Bryan} \& {Norman}(1998)}]{bryan98}
{Bryan}, G.~L. \& {Norman}, M.~L. 1998, \apj, 495, 80

\bibitem[{{Bullock} {et~al.}(2001){Bullock}, {Kolatt}, {Sigad}, {Somerville},
  {Kravtsov}, {Klypin}, {Primack}, \& {Dekel}}]{bullock01}
{Bullock}, J.~S., {Kolatt}, T.~S., {Sigad}, Y., {Somerville}, R.~S.,
  {Kravtsov}, A.~V., {Klypin}, A.~A., {Primack}, J.~R., \& {Dekel}, A. 2001,
  \mnras, 321, 559

\bibitem[{{Buote} {et~al.}(2006){Buote}, {Gastaldello}, {Humphrey},
  {Zappacosta}, {Bullock}, {Brighenti}, \& {Mathews}}]{buote06b}
{Buote}, D.~A., {Gastaldello}, F., {Humphrey}, P.~J., {Zappacosta}, L.,
  {Bullock}, J., {Brighenti}, F., \& {Mathews}, W. 2006, \apj, submitted,
  (astro-ph/0610135)

\bibitem[{{Buote} {et~al.}(2003){Buote}, {Lewis}, {Brighenti}, \&
  {Mathews}}]{buot03a}
{Buote}, D.~A., {Lewis}, A.~D., {Brighenti}, F., \& {Mathews}, W.~G. 2003,
  \apj, 594, 741

\bibitem[{{Cavaliere} \& {Fusco-Femiano}(1978)}]{beta}
{Cavaliere}, A. \& {Fusco-Femiano}, R. 1978, \aap, 70, 677

\bibitem[{{Dickey} \& {Lockman}(1990)}]{dick90}
{Dickey}, J.~M. \& {Lockman}, F.~J. 1990, \araa, 28, 215

\bibitem[{{D'Onghia} {et~al.}(2005){D'Onghia}, {Sommer-Larsen}, {Romeo},
  {Burkert}, {Pedersen}, {Portinari}, \& {Rasmussen}}]{donghia05}
{D'Onghia}, E., {Sommer-Larsen}, J., {Romeo}, A.~D., {Burkert}, A., {Pedersen},
  K., {Portinari}, L., \& {Rasmussen}, J. 2005, \apjl, 630, L109

\bibitem[{{Ettori}(2000)}]{ettori00}
{Ettori}, S. 2000, \mnras, 311, 313

\bibitem[{{Ferguson} \& {Sandage}(1990)}]{ferguson90}
{Ferguson}, H.~C. \& {Sandage}, A. 1990, \aj, 100, 1

\bibitem[{{Fukugita} {et~al.}(1995){Fukugita}, {Shimasaku}, \&
  {Ichikawa}}]{fukugita95}
{Fukugita}, M., {Shimasaku}, K., \& {Ichikawa}, T. 1995, \pasp, 107, 945

\bibitem[{{Gastaldello} {et~al.}(2006){Gastaldello}, {Buote}, {Humphrey},
  {Zappacosta}, {Bullock}, {Brighenti}, \& {Mathews}}]{gast06a}
{Gastaldello}, F., {Buote}, D.~A., {Humphrey}, P.~J., {Zappacosta}, L.,
  {Bullock}, J.~S., {Brighenti}, F., \& {Mathews}, W.~G. 2006, ArXiv
  Astrophysics e-prints, \apj, submitted (astro-ph/0610134)

\bibitem[{{Grevesse} \& {Sauval}(1998)}]{grsa98}
{Grevesse}, N. \& {Sauval}, A.~J. 1998, Space Science Reviews, 85, 161

\bibitem[{{Jeltema} {et~al.}(2006){Jeltema}, {Mulchaey}, {Lubin}, {Rosati}, \&
  {B{\"o}hringer}}]{jeltema06}
{Jeltema}, T.~E., {Mulchaey}, J.~S., {Lubin}, L.~M., {Rosati}, P., \&
  {B{\"o}hringer}, H. 2006, \apj, 649, 649

\bibitem[{{Jones} {et~al.}(2003){Jones}, {Ponman}, {Horton}, {Babul},
  {Ebeling}, \& {Burke}}]{jones03}
{Jones}, L.~R., {Ponman}, T.~J., {Horton}, A., {Babul}, A., {Ebeling}, H., \&
  {Burke}, D.~J. 2003, \mnras, 343, 627

\bibitem[{{Khosroshahi} {et~al.}(2006){Khosroshahi}, {Ponman}, \&
  {Jones}}]{khosroshahi06}
{Khosroshahi}, H.~G., {Ponman}, T.~J., \& {Jones}, L.~R. 2006, \mnras, 372, L68

\bibitem[{{Kinney} {et~al.}(1996){Kinney}, {Calzetti}, {Bohlin}, {McQuade},
  {Storchi-Bergmann}, \& {Schmitt}}]{kinney96}
{Kinney}, A.~L., {Calzetti}, D., {Bohlin}, R.~C., {McQuade}, K.,
  {Storchi-Bergmann}, T., \& {Schmitt}, H.~R. 1996, \apj, 467, 38

\bibitem[{{Landolt}(1992)}]{landolt92}
{Landolt}, A.~U. 1992, \aj, 104, 340

\bibitem[{{Markevitch}(1998)}]{mark98b}
{Markevitch}, M. 1998, \apj, 504, 27

\bibitem[{{Mathews} {et~al.}(2005){Mathews}, {Faltenbacher}, {Brighenti}, \&
  {Buote}}]{mathews05}
{Mathews}, W.~G., {Faltenbacher}, A., {Brighenti}, F., \& {Buote}, D.~A. 2005,
  \apjl, 634, L137

\bibitem[{{Milosavljevi{\'c}} {et~al.}(2006){Milosavljevi{\'c}}, {Miller},
  {Furlanetto}, \& {Cooray}}]{milos06}
{Milosavljevi{\'c}}, M., {Miller}, C.~J., {Furlanetto}, S.~R., \& {Cooray}, A.
  2006, \apjl, 637, L9

\bibitem[{{Mulchaey} {et~al.}(2006){Mulchaey}, {Lubin}, {Fassnacht}, {Rosati},
  \& {Jeltema}}]{mulch06}
{Mulchaey}, J.~S., {Lubin}, L.~M., {Fassnacht}, C., {Rosati}, P., \& {Jeltema},
  T.~E. 2006, \apj, 646, 133

\bibitem[{{Nelson} {et~al.}(2002){Nelson}, {Gonzalez}, {Zaritsky}, \&
  {Dalcanton}}]{nelson02}
{Nelson}, A.~E., {Gonzalez}, A.~H., {Zaritsky}, D., \& {Dalcanton}, J.~J. 2002,
  \apj, 566, 103

\bibitem[{{Osmond} \& {Ponman}(2004)}]{osmond04}
{Osmond}, J.~P.~F. \& {Ponman}, T.~J. 2004, \mnras, 350, 1511

\bibitem[{{Pierre} {et~al.}(2004){Pierre}, {Valtchanov}, {Altieri}, {Andreon},
  {Bolzonella}, {Bremer}, {Disseau}, {Dos Santos}, {Gandhi}, {Jean}, {Pacaud},
  {Read}, {Refregier}, {Willis}, {Adami}, {Alloin}, {Birkinshaw}, {Chiappetti},
  {Cohen}, {Detal}, {Duc}, {Gosset}, {Hjorth}, {Jones}, {LeFevre}, {Lonsdale},
  {Maccagni}, {Mazure}, {McBreen}, {McCracken}, {Mellier}, {Ponman},
  {Quintana}, {Rottgering}, {Smette}, {Surdej}, {Starck}, {Vigroux}, \&
  {White}}]{pierre04}
{Pierre}, M., {et~al.} 2004, Journal of Cosmology and Astro-Particle Physics,
  9, 11

\bibitem[{{Saxton} \& {Siddiqui}(2002)}]{saxton02}
{Saxton}, R.~D. \& {Siddiqui}, H. 2002, XMM-SOC-PS-TN-43

\bibitem[{{Sheinis} {et~al.}(2002){Sheinis}, {Bolte}, {Epps}, {Kibrick},
  {Miller}, {Radovan}, {Bigelow}, \& {Sutin}}]{sheinis02}
{Sheinis}, A.~I., {Bolte}, M., {Epps}, H.~W., {Kibrick}, R.~I., {Miller},
  J.~S., {Radovan}, M.~V., {Bigelow}, B.~C., \& {Sutin}, B.~M. 2002, \pasp,
  114, 851

\bibitem[{{Suto} {et~al.}(1998){Suto}, {Sasaki}, \& {Makino}}]{suto98}
{Suto}, Y., {Sasaki}, S., \& {Makino}, N. 1998, \apj, 509, 544

\bibitem[{{Voit}(2005)}]{voit05b}
{Voit}, G.~M. 2005, Rev. Mod. Phys., 77, 207

\bibitem[{{White} {et~al.}(1994){White}, {Giommi}, \& {Angelini}}]{white94}
{White}, N.~E., {Giommi}, P., \& {Angelini}, L. 1994, \iaucirc, 6100, 1

\bibitem[{{Willis} {et~al.}(2005){Willis}, {Pacaud}, {Valtchanov}, {Pierre},
  {Ponman}, {Read}, {Andreon}, {Altieri}, {Quintana}, {Dos Santos},
  {Birkinshaw}, {Bremer}, {Duc}, {Galaz}, {Gosset}, {Jones}, \&
  {Surdej}}]{willis05}
{Willis}, J.~P., {et~al.} 2005, \mnras, 363, 675

\bibitem[{{Zentner} {et~al.}(2005){Zentner}, {Berlind}, {Bullock}, {Kravtsov},
  \& {Wechsler}}]{zentner05}
{Zentner}, A.~R., {Berlind}, A.~A., {Bullock}, J.~S., {Kravtsov}, A.~V., \&
  {Wechsler}, R.~H. 2005, \apj, 624, 505

\end{thebibliography}
\end{document}